# On the Categorization of Nanomaterials

**John Rumble [a,b]**

[a] R&R Data Services, Gaithersburg MD, USA
[b] CODATA Working Group on Nanomaterials, Paris, France

Email: john.rumble@randrdata.com

ORCID iD: 0000-0001-6705-5768

**Abstract**

In this paper, categorization of nanomaterials is examined from four perspectives – context, criteria for success, ensuring measurements are relevant, and the life cycle of a nanomaterial. For each perspective, its relevance to categorization is discussed as well as the difficulties it presents. For example, while the context of assessing potential harm to living things and the environment is clearly important, other contexts are often needed and require different categorization schemes. Understanding what success means for a categorization scheme, within its target context, is critical to making sure a categorization is actually useful. The complexity of nanomaterials and their interactions makes generating and collecting the required data and metadata to support categorization a challenge. Finally, the transformation a nanomaterial undergoes through its lifetime, including the testing process, present additional challenges to accurate categorization. How these factors impact development of usable categorization schemes is analyzed.

**Keywords**
Nanomaterial complexity, categorization of nanomaterials, nanostructured materials, nanoscale context

**Introduction**

The wide variety, diversity, and reactivity of nanomaterials preclude systematic measurement of all properties under all conditions; consequently, the need to predict properties relevant to specific conditions and uses is quite important. While predictive models exist for a small number of nanomaterials [1], those models are limited to a few specific circumstances and properties. Based on the success of Quantitative Structure-Activity Relationships (QSARs) methods in chemistry and drug design [2], grouping nanomaterials into categories is important so that properties of representative class members can be used to predict properties of other class members.

To date, the majority of efforts related to nanomaterials categorization have focused on two contexts: (1) registration of a commercially produced chemical in the European Union and (2) assessment of the potential harmful effects of nanomaterials on living things and the environment. Successful categorization in these contexts can greatly facilitate the acceptance, use, and commercialization of nanomaterials. While considerable progress has been made, there are broader issues with respect to categorization that have not yet been fully explored that impact not only these two contexts, but also categorization in general. In this paper, the categorization of nanomaterials is discussed from four perspectives that reflect their complexity as well as their potential benefits but have not yet been fully addressed. Each perspective is discussed below in more detail.

- Categorization context
- Criteria for success of a categorization approach
- Ensuring measurements are relevant to categorization
- Impact of the life cycle of a nanomaterial.

**Categorization context**
Categorization is useful in many different contexts, and each context can result in a different categorization approach. Though nanomaterial categorization for purposes of helping to understand the potential toxicity has emerged as primary goal, other contexts are important with toxicity. Here we will discuss four contexts of importance to nanomaterials. First, however, it must be pointed out that each categorization context emphasizes a different set of properties and features. It is unlikely that the same set of properties is important to multiple contexts. A second point to consider in looking at a context is that it is assumed that each member of a category has property values for the classifying properties that lie within a specified range. Consequently, when comparing two different categorization schemes, those property value ranges may not overlap. Therefore, the predictive power of one category scheme does not carry over to another.

*Fitness of purpose*
Categorization related to fitness of purpose is critical with respect to the design of new or improved nanomaterials. Fitness for purpose covers a wide range of uses, including the efficacy of drug delivery, ease of post-manufacturing processing such as coatings, suitability for inclusion in bulk materials, electronic properties of quantum dots, etc. Fitness for purpose categorization is extremely important in commercial situations. Companies are constantly trying to develop new and improved products, and experience with traditional materials has shown that many *new* materials result from major or minor changes to existing materials to enhance one or more desired properties or features. This experience has also shown that having well-defined categories of existing materials based on a well-established understanding of cause and effect (e.g. the electronic properties of alloys used in hard disks [3] supports directed R&D with specific goals (fitness of purpose). The same applies for bio-organic materials such as pharmaceuticals, where a clear understanding of the active features of a drug shapes investigation of more effective variants. Fitness of purpose categorization is also extremely important for users of nanomaterials, who are equally interested in finding substitute, cheaper, or more effective materials to use in existing products.

Most major companies using materials and chemicals have developed internal categorization schemes for these items focused on properties and features of interest to them, e.g., color, durability, etc. A major point to recognize is that categorizing nanomaterials for fitness of purpose creates groupings reflecting an understanding of the features affecting a specific purpose that may not correlate with other groupings such as for potential toxicity.

*Environmental accumulation*
The accumulation of nanomaterials in the environment has been studied aggressively (see for example references [4-6]) over the last decade, with the recognition that the size and surface reactivity of nanomaterials can greatly facilitate accumulation. From the viewpoint of the life cycle of a nanomaterial, accumulation occurs after many earlier steps, and categorization must take into account the variety of coatings or dissolution that a nanomaterial has experienced. It is not clear that the features relevant to grouping for accumulation activity have been well identified (see for example reference [7]) or correlated with other grouping features.

*Potential toxicity*
The potential for nanomaterials to do harm to living things has been the biggest driver for categorization efforts to date. The ability to predict the potential for harm (toxicity) without extensive testing is a major goal of the nanomaterials community and a critical factor to acceptance by regulatory agencies and commercial success.

Extensive work on the toxicity of nanomaterials has already been published, far more than can be reviewed in detail here. (See for example references [8, 9]). Much of it has been done without regard to realistic exposures and dosages [10], but valuable nonetheless for pointing out areas of possible concern. In addition, considerable effort has gone into developing predictive techniques based on categorization criteria, which appear to have much applicability, though lacking in enough quality data to provide assurance of usefulness [11-15]. One concern is the poor characterization of the nanomaterial being tested, which is discussed below in further detail. Another concern is the lack of causal mechanisms for many toxicity endpoints that make the predictions based more on heuristics than experimental evidence for cause and effect. Regardless of present-day shortcomings, the assessment of potential toxicity is a major context and force driving the development of useful categorization schemes.

*Exposure (Availability)*
The impact of nanomaterials on an organism is a function of the toxicity and exposure (including availability and dosage) experienced by the organism. A substance may have harmful effects on living things, but if it is not available (that is, it is not exposed, to the organism), the potential harm may be mitigated. The exposure, which is the amount of a nanomaterial that is available to affect an organism by accumulation, is highly dependent on the rate of dissolution, reactivity, surface coatings, and other factors. Categorizing nanomaterials with respect to their ease of availability has not been well studied to date (see for example reference [4]) yet remains an important context. Because nanomaterials in use are often embedded in a solid or liquid media, system considerations (nanomaterials plus media) are very important. Given the number of different nanomaterials and the number of different possible media, the number of systems to be considered and categorized is quite large.

One additional exposure scenario of importance is worker exposure, including inhalation processes. They are likely to be exposed to nanomaterials either during manufacture and distribution or being processed into a product.

*Cross-context relevancy*

As pointed out above, the various contexts that one can consider for nanomaterial categorization can interact and cross-contextual categorization is a major concern. What that means is that the independent variables that underly cause-and-effect in one context may not be relevant in another context. This situation can arise from several circumstances. As a result, data gathered for one context may not be useful in another context as the independent variables being emphasized in the measurements may not illuminate cause-and-effect in another context. Further, even if relevant, the data may have not been included in the publication.

**Criteria for success of a categorization approach**

How does one determine if a categorization scheme is successful, that is, useful and usable for its intended purpose? Having stated criteria for success provides guidance not only for applicability (predictive capability) but also for extendibility. This section puts forth criteria for assessing the suitability and success of a categorization approach. These criteria have been developed for applications not related to nanomaterials and have been adapted for nanomaterial contexts.

The general purpose of a categorization system is to group a subset of individual items in an entire population of items of interest such that each item in a class has the same (within a range of values) set of characteristics, properties, or functionality as defined by the class definitions. Consequently, if an item is identified as a member of a class, then it can be assumed to have characteristics, properties, and functionalities that lie within the boundaries defined as by the class.

Before looking at possible criteria for determining success, it is important to review the assumptions normally made when categorizing nanomaterials. These assumptions restate many premises of good scientific practice.

1. Multiple categorization approaches for nanomaterials are possible, even within one context
    Nanomaterials can be classified using different characteristics, properties and functionalities, such as size, shape, major chemical component, toxicity, cost, surface reactivity, and many others, resulting in many different possible groupings.

2. Nanomaterial categorization systems are useful in many different contexts, with considerable importance presently given for the prediction of biological and environmental impacts.

These impacts can be positive, negative, neutral, and indeterminate, but their predictability is a key consideration in commercial activity.

3. A nanomaterial categorization scheme should ideally be based on a *few* properties or features that have been *clearly demonstrated*, if possible, to have a *cause-and-effect relationship* with important end-points.

    These features and properties should have been demonstrated, to the extent possible, to have a cause-and-effect relationship with the specific biological and environmental impact that are desired or are to be avoided.

4. The features and properties used for categorization must be measurable quantitatively and be reproducible by well-documented, and to the extent practicable, standardized measurement techniques

    The measurement technology used must be well researched and adopted by the community after rigorous testing so that an individual measurement with a stated uncertainty can be accepted as accurate.

5. A causality model should link the categorization feature or property with the outcome (impact); correlation and heuristic models are not causality; one must be able to control feature to demonstrate causality.

Hill [16] and Rothman and Greenland [17] presented several criteria for differentiating between causality and association as summarized below. The criteria, as adapted below, were developed in the field of epidemiology but are directly applicable to the assessment of the quality, i.e., the *success*, of a nanomaterial categorization system. Notes have been added identifying how each criterion is relevant to nanomaterial categorization, especially is assessing the validity and success of a categorization scheme.

- *Consistency*: Consistent measurements made by different experimenters in different places with different samples using documented procedures strengthen the likelihood of an effect

    Properties of a substance (here a nanomaterial) are established over time and require many repeated measurements that demonstrate that all important independent variables have been identified and controlled. Rarely does one measurement definitely establish a property value. Not only is consistency necessary to establish a value, and by definition, reproducibility, but also is required to define a meaningful uncertainty.

- *Specificity*: Causation is likely if a specific measurement result is shown to be dependent on well-identified independent variable(s) with no other likely explanation. The more specific an association between a variable(s) and an effect is, the bigger the probability of a causal relationship.

    A wide variety of independent variables govern the behavior and interactions of a nanomaterial (size, shape, chemical composition, physical structure, surface structure and charge, coatings, etc.) so that care must be taken to establish a causal relationship between each independent variable and an effect. Moreover, the possibility of co-factors, that is, multiple variables synergistically causing a result, must carefully be ruled out. Presently ISO, ASTM, and OECD have established

measurement protocols for some properties that increase their specificity of cause-and-effect but given the variety of systems (nanomaterial plus bulk material or liquid media) that constitute real-life exposure of nanomaterials, that specificity is sometimes suspected.

- *Biological gradient*: Greater exposure should generally lead to greater incidence of the effect. However, in some cases, the mere presence of the factor can trigger the effect.

  Ensuring that biological exposure during testing is similar to that experience in real-life situations is a vexing problem for nanomaterials. As Krug has pointed out [10], much testing has been done at very high doses or at a single dose. In the first case, the dosages may be unrealistically high, and results reflect a response based on overwhelming dosage. In the latter case, single dose testing does not establish a meaningful biological gradient, masking whether there is a lower-limit threshold or whether the response is linear or non-linear. In addition, biological testing needs to be carefully controlled with respect to coatings, which are ubiquitous for nanomaterials, dissolution rates, transformations such as agglomeration and aggregation. Classical toxicity testing is designed to take into account such issues, but nanomaterials add another level of complexity due to their reactivity.

- *Strength*: A small association does not mean that there is not a causal effect, though the larger the association, the more likely that it is causal.

  Strength is closely associated with biological gradient, though not identical. Strength refers to the strength of a correlation between a feature and an effect but does not establish actual cause and effect. Dissolution rate might strongly correlate to cellular take-up of a nanomaterial. Dissolution itself, however, is not a cause of toxicity. As another example, as doses of a poison are increased, the harmful effects obviously increase, from impairment of motion to actual death.

- *Plausibility*: A plausible mechanism between cause and effect is critical; Hill noted that understanding of the mechanism may be limited by current knowledge [16].

  Establishing cause-and-effect requires demonstration that the effect never occurs without the cause and that controlling the cause can lead to predictable effects. That, however, does not always discover the actual mechanism linking the cause to the effect. While it is not necessary to know the specific mechanism that makes something harmful in recognizing that harm occurs – for example, ingesting rat poison can clearly show a harmful effect (death) to the average person without having to know how the poison acted, but in ascribing a harmful effect to a feature or a combination of features (e.g. specific surface areas and surface chemistry) of a nanomaterial to build a categorization scheme requires some understanding of the mechanism so that other members of the proposed category can be classified with some assurance of accuracy. This is especially true when the deleterious effect may not be totally ascribed to a single feature or characteristic.

- *Coherence*: Coherence between theoretical and laboratory findings increases the likelihood of an effect.

Two types of models are of importance in nanomaterials categorization: one for the prediction of properties and functionalities and the other that describes the state of a nanomaterial (nanoform or nano-object – see below) at a specific moment. Computation models that predict nanomaterial properties, especially from first principles, are limited today but growing in number, especially for complex behavior. For example, nanoHUB (nanoHUB.org) has over 320 simulation tools available for electronic, mechanical, biological, and other nanomaterial applications.

The physical models that are built to describe the life cycle of a nanomaterial, from its manufacture to its use to its effect in a biological system are an abstraction of a very complex set of transformations that are not easily quantified through detailed measurements on a step-by-step basis. It is reasonable to assume that if the "starting" material for a biological effect test is well characterized, then how it got to that state will not be important for establishing cause-and-effect. That said it is less reasonable to assume a group of nano-objects undergoing the same life cycles (transformations) will end up as identical items. Then it becomes important to ascertain the "range" of parameters that occur within the group and how that range might affect measurement results. The simplest example, of course, is the size distribution, but many other features experience the spread of values. Computation models that predict such ranges, and the consequent range of test results, need to support the required coherence. For example, the different test methods required to determine size distributions within different ranges do not always mesh well with the theoretical predictions of the computational models [18, 19].

In terms of assessing the validity of a categorization scheme, coherence can be often overlooked. The key point here is that because a categorization scheme assumes all members have certain properties, characteristics, or functionalities that lie within a specified range, the models used to predict behavior must be applicable to each member of the category.

- *Analogy*: The effect of similar factors may be considered.

The complexity of nanomaterials and their interactions combined with the large number of interesting nanomaterials being researched and produced is a significant barrier to comprehensive testing. Indeed, this fact is a primary motivation for developing useful categorization schemes so identification of a category to which a new or novel nanomaterial can be assigned membership is critical. Assigning property value ranges by analogy through structure-activity relationships or "Read-Across" [13] is very important and a subject of much present-day work. What is equally important is that assignment by analogy can be verified when needed – almost on an "On-demand" basis.

What this means with respect to the assessment of the quality of a categorization scheme is that periodic review of predictions made by analogy – by simple extrapolation or more complex models – must be rigorously pursued. Over time, it is possible that the "allowable" deviations from the range of acceptable property values can drift, resulting in category membership that "almost fits, but not quite." This defeats the purpose of categorization but is to be expected. Often when this happens, the category boundaries need to be reassessed. This is not to be decried on basic principles; indeed, when the context of categorization is to discover "fitness for purpose," such expansion can open up opportunities to explore new materials or novel properties. Such expansion is in fact the basis of innovation. In the context of predicting unwanted effects, the relaxation of a parameter range may introduce nanomaterials with unwanted effects.

- *Temporality*: The effect must occur after the cause; if there is an expected delay between the cause and expected effect, then the effect must occur after that delay.

  Temporality is perhaps less important as an assessment tool for categorization of nanomaterials than for an area such as epidemiology or disease exposure. At the same time, temporality does play a role when looking at nanomaterial life cycle steps and ensuring that interactions might be mitigated by coatings, coronas, aggregation, and agglomeration unless those transformations are correctly placed temporally in a cause and effect chain.

Given the growing importance of categorization in the context of toxicity and environmental concentration, *assessment* of the validity of a specific categorization scheme is necessary. The criteria enumerated above provide a useful framework for such assessments, especially with respect to identifying critical cause-and-effect linkage. The potential commercial value of nanomaterials is extremely high and excluding a useful nanomaterial erroneously because of poor categorization not only costs money but also deprives society of a potentially valuable nanomaterial.

**Ensuring measurements are relevant to categorization**

As pointed out in this special collection, one of the largest barriers to establishing a consensus-driven, harmonized categorization system for nanomaterials is the lack of high quality data upon which causality can be established. There are four basic steps to developing a database suitable for categorization:
- Defining meaningful property(s) or endpoint(s) for categorization
- Establishing rigorous test methods for the property(s) or endpoint(s)
- Developing data and metadata reporting standards so that causality can be established, verified, and documented
- Building and disseminating data repositories and databases of high quality measurement results

*Defining meaningful property(s) or endpoint(s) for categorization*
Each categorization context defines one or more criteria for inclusion in its scheme. The suitability of these criteria is not be discussed here except to point out that they should be clear, tied to a measurable property or feature, and to the extent possible, have both their cause and effect and mechanism of action well understood. Over time the criteria for membership can and will change as new knowledge is developed

For regulatory purposes, the need for clarity and relevance is especially important because of commercial considerations. If the property or endpoint is not significant, time and money is wasted on optimizing to incorrect criteria. Here balance must be maintained between "It would be nice to have this measurement result" and "It is necessary to have this measurement result". To avoid the former, having a well-documented cause-and-effect mechanism is crucial. To date, the emphasis has been on avoiding potential problems, but that approach over time may change, for example, if in certain occupational settings other evidence shows that nanomaterials themselves are generally not harmful because of their size. There is quite a bit of data on the inhalation toxicity of materials that are respirable (less than 10 microns) and poorly soluble, especially at relatively low dose levels. Endpoints that separate inhalation from other exposure pathways, e.g., dermal, oral, and degradation of matrices containing nanomaterials, may be needed.

*Establishing rigorous test methods for the property(s) or endpoint(s)*
Internationally recognized standards development organizations, such as ISO TC 226 Nanotechnology [20] and ASTM E56 on Nanotechnology [21], as well as OECD [22] have made considerable progress in setting up well-documented and rigorous tech procedures for important property measurements. These methods have been well documented and are not discussed in detail here. It suffices to say that the development of such procedures is time-consuming and costly so that the number of adopted procedures is increasing slowly.

*Developing data and metadata reporting standards so that causality can be established, verified, and documented*
Prior to the Information Age and the emergence of databases – bibliographic and numerical, there were virtually no standards or set procedures for reporting research results. Today, when virtually all information is produced digitally, and more importantly, shared digitally, standards for reporting the written and factual scientific record are not only important, but also almost mandatory because sharing information and data is routine and expected. Traditionally, these formal and informal standards have been used for several purposes.

- Documenting what has been done
- Allowing reproduction of measurement and results
- Validating that all independent variables have been identified and controlled
- Establishing uncertainty
- Providing insights for new investigations

In the case of nanomaterials and the importance of categorization for regulatory purposes, an additional purpose for data and metadata standards is to assure that enough sufficient data and metadata are available for assessment of potential negative effects, which in turn supports categorization for that context. These subjects have also been the subject of a series of papers on nanoinformatics written by the U.S. National Cancer Institute Nanoinformatics Working Group [23-26].

Building data repositories of nanomaterials property data, such as eNanoMapper [27], the Nanomaterials Registry [28], CEINT [29], CEIN [30], and others, implicitly develop informal standards, but these standards are applied *after* the data and metadata have been reported. To date, publications do not require authors report data and metadata in a manner directly usable by such databases, in contrast to the requirements for crystallographic result publication [31]. Several regions and countries have requirements for reporting *standardized* nanomaterials metadata including Europe [32] and the United States (see for example requirements by different U.S. Agencies [33]).

There is a more subtle reason for creating these standards, and that has to do with the need to have enough information to establish cause-and-effect relationships. The number of independent variables that contribute to the interactions and reactions of nanomaterials is quite large. Aside from the informal standards mentioned above, other metadata reporting systems are those contained in ISA-TAB-nano [34] and the OECD test guidelines [35,36]. A recent systematic attempt to define the variables important for characterizing nanomaterials and provide guidance on how to report them is that of the CODATA Uniform Description System (UDS) [37,38]. The UDS defines a *nano-object* as the smallest meaningful unit of a nanomaterial and enumerates seven *information categories* for data and metadata that can be used to report quantitative information to allow a nano-object to be described uniquely. Other critical aspects of nanomaterials such as surfaces, interfaces, topology, and morphology are not yet understood well enough to define the independent variables necessary to quantify those aspects.

The UDS can be used to develop ontologies, set up and validate database schemas, or to define reporting requirements for publications and regulatory actions. Recently three ASTM Standard Guide have been approved based on the UDS: one for reporting the physical and chemical characteristics of nano-objects (E3144), one for reporting of the physical and chemical characteristics of a collection of nano-objects (E3) and one for reporting production information and data for nanomaterials under the auspices of ASTM E56 [39].

REACH and EPA have introduced the concept of *nanoform*, as a basic unit of interest [40, 41]. The definitions of nanoform are still ambiguous, and it is difficult to use those definitions to establish cause-and-effect. It should be noted that the UDS as described above can be used to make the description of a nanoform much more precise. The most comprehensive ontology for nanoparticles was developed several years ago [42], and major portions of it have been included in ISA-TAB-Nano and its extensions [34]. OECD templates for reporting results of various standardized nanomaterial tests have very limited description of the tested nanomaterial, which severely limits establishing cause-and-effect accurately [35]. Clearly additional work is needed to integrate these efforts into a cohesive and comprehensive set of standards so accurate descriptions of nanomaterials and test results can be agreed upon internationally.

This brief review of existing approaches to describing nanomaterials with enough detail to ascribe cause-and-effect is given to demonstrate our lack of knowledge of mechanisms of action for nanomaterials, which is to be expected given the complex nature of nanomaterials. Teasing out the details of which specific features of a nanomaterial are the cause of a specific effect takes time and many years of research. Future years should show considerable progress in gaining more knowledge of how nanomaterials actually interact with biological systems.

The implication for categorization is quite clear. Until we know the details of the mechanisms of actions, ascribing an effect to one or more specific features is difficult and grouping nanomaterial on the basis of common important features equally difficult. Context again is also important because different endpoints may manifest differently to subtle changes in the features.

*Building and disseminating data repositories and databases of high quality measurement results*
Categorization requires sufficient amounts of reliable data to be done accurately. Quantifying a sufficient amount is difficult , but the large databases developed for QSAR work in the pharma industry were the result of decades of work generating data on hundreds of thousands of compounds with hundreds of properties reported. This amount of data is not yet available for nanomaterials, though high-throughput screening techniques can possibly fill gaps more quickly than the individualized experimental measurements made in the latter half of the twentieth century. The few databases of nanomaterials properties available today are just beginning to build useful data volumes.

A slightly different approach is possible, which is to build a complete characterization of a specific set of nanomaterials that would enable discerning cause-and-effect for a specific situation and use the results to build limited but still useful categories. This seems to be the reasoning behind the concept of *nanoforms*, but its long-term viability to predict critical behavior remains to be proven.

To summarize the present day situation with respect to ensuring measurements are relevant to categorization, many positive steps have been taken to build the body of data and knowledge needed to put nanomaterial categorization in several contexts on a solid scientific foundation. As pointed out above, much work remains.

**Life cycle of nanomaterials**
Nanomaterials are very reactive and are subject to several deliberate and random changes (transformations) over their life cycle. For example, many commercial nanomaterials are coated to enhance a property (such as dispersibility), and virtually all nanomaterials pick up one or more coatings when exposed to a liquid media. The transformations experienced by a nanomaterial during its life cycle have considerable impact on categorization efforts. Some of these impacts and their consequences on categorization are discussed next. The impacts fall into three types.

- Uncertainty of the nature of the nanomaterial being tested
- Uncertainty of cause and effect
- Uncertainty with respect to categorization

*Uncertainty of the nature of the nanomaterial being tested*
Figure 1 outlines the typical sequence of events involved in testing a nanomaterial. During the testing process, a nanomaterial usually goes through one or more transformations, and unless each step of the process is carefully controlled, the relationship of the test result to the *original* (manufactured, natural, produced) nanomaterial is not always clear. Further, the features of the nanomaterial that govern the test result (cause and effect) may not always be traceable to specific features of the original nanomaterial. With care, of course, these problems can be overcome, especially by careful characterization of the nanomaterial at each step of the process. A recent ASTM standard guide on the production of nano-objects can be helpful [39]

*Uncertainty of cause and effect*
Because the changes that often take place through the life cycle can substantially impact features such as size (through agglomeration and aggregation), surface chemistry (coatings, coronas, etc.), surface charge distribution (through changes in shape), ascribing cause and effect to a feature on the original nanomaterial can be difficult. Specifically, transformations are dependent on the exposure history and environment, and subtle changes in these can significantly change the transformation outcome. This in turn complicates the assignment of cause and effect to original features.

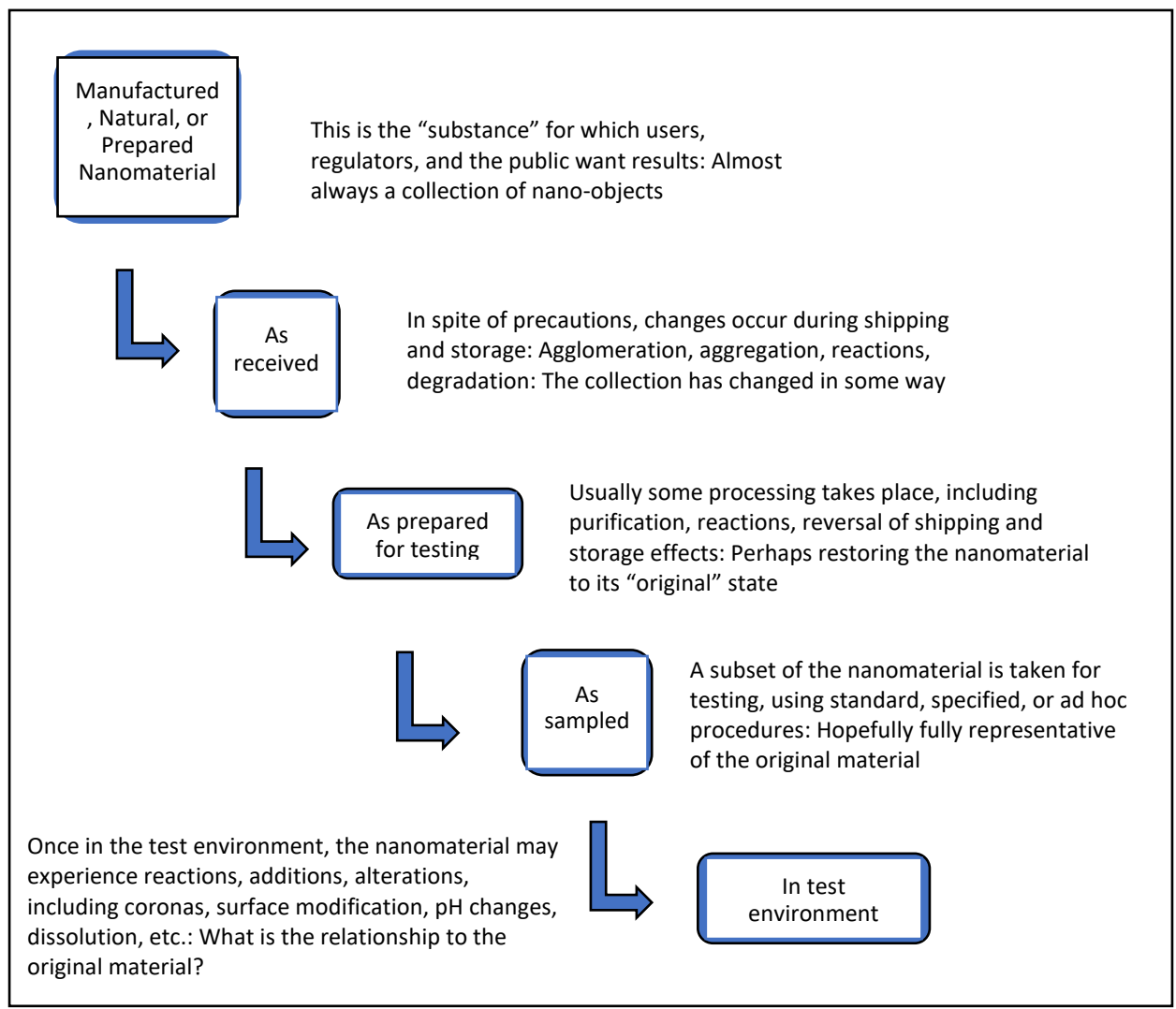

**Figure 1:** Schematic of life cycle steps involved in testing a nanomaterial

*Uncertainty with respect to categorization*

It is easy to say that categorization of nanomaterials should be based on what is produced (upper left box in Figure 1), as that is the substance that needs to be registered with various country and regional regulatory authorities. It should be clear, however, that there is considerable ambiguity to the meaning of categorization on that basis. Questions that arise include the following:

- Is a test result solely the function of the original nanomaterial and its features, or dependent also on one or more steps in its life cycle?

- If a nanomaterial, more specifically one nanoform as defined by national or regional regulatory agencies, is changed, does that changed nanomaterial (nanoform) go through its life cycle differently or the same as the original nanoform?

- Finally, and most importantly, does the dependency of the test results on life cycle history preclude accurate prediction of the properties of other members of the category because the mechanism of action may be altered?

The answers to these questions require more knowledge than we have at present, but new programs such as GRACIOUS in the EU [43] are being designed to explore this complexity.

**Integrating the four perspectives into a world-view of categorization**

In this paper, we have looked at categorization of nanomaterials from four perspectives.

- Categorization context
- Criteria for success of a categorization approach
- Ensuring measurements are relevant to categorization:
- Impact of the life cycle of a nanomaterial:

The intent has not been to say that categorization is wrong or impossible; instead it has been to take the complexity of nanomaterials as has been unfolded through years of research and use that knowledge to refine our concept of what categorization should be and how can we proceed on a scientifically correct basis. The initial premise of categorization remains unchallenged: that through the grouping of similar nanomaterials based on scientifically sound principles we can overcome the problem of the impossibility of testing every nanomaterial and every variant thereof for every possible property. Without categorization success, this problem is insurmountable, high-throughput screening notwithstanding.

It is hoped that by confronting the issues raised by the examination of these four perspectives and the insights provided by examining these perspectives thoroughly, researchers will find answers to the questions raised and solutions to the problems identified. The value to society of nanomaterials, especially the idea that today we can build a substance on the nanoscale to have the properties and functionalities that can make our lives better, is so high that we must make categorization a valuable tool for the future.

**Acknowledgements**
The author acknowledges support from the FutureNanoNeeds project (FP7-NMP-2013-LARGE-7) and from CODATA, Paris France. The author wishes to thank Maria Doa and Betul Hekimoglu-Balkan for clarifying discussions.

This article is one of a collection of articles about the categorization of nanomaterials, generated by research and workshop discussions under the FutureNanoNeeds project funded by the European Union Seventh Framework Programme (Grant Agreement No 604602). For an overview and references to other articles in this collection, see *The Nature of Complexity in the Biology of the Engineered Nanoscale Using Categorization as a Tool for Intelligent Development* by Kenneth A. Dawson.

Author declares that he has no conflict of interest.